\documentclass[aps,twocolumn,amsmath,amssymb,showpacs,prb,superscriptaddress,unsortedaddress]{revtex4}

\usepackage{graphicx}
\usepackage{verbatim}
\usepackage{mathrsfs}
\usepackage{sidecap}

\usepackage{amsmath,amsfonts,amssymb}
\usepackage{graphicx}

\def\3{2.8in}    
\def\2{2.5in}
\def\4{3.0in}

\def \beq {\begin{equation}}
\def \eeq {\end{equation}}
\begin{document}

\title{Topological phase diagram and saddle point singularity in a tunable topological crystalline insulator}






\author{Madhab Neupane}\affiliation {Joseph Henry Laboratory, Department of Physics, Princeton University, Princeton, New Jersey 08544, USA}
\affiliation {Condensed Matter and Magnet Science Group, Los Alamos National Laboratory, Los Alamos, NM 87545, USA}


\author{Su-Yang Xu}\affiliation {Joseph Henry Laboratory, Department of Physics, Princeton University, Princeton, New Jersey 08544, USA}

\author{R. Sankar} \affiliation{Center for Condensed Matter Sciences, National Taiwan University, Taipei 10617, Taiwan}

\author{Q. Gibson}\affiliation {Department of Chemistry, Princeton University, Princeton, New Jersey 08544, USA}

\author{Y. J. Wang}\affiliation {Department of Physics, Northeastern University, Boston, Massachusetts 02115, USA}
\affiliation {Advanced Light Source, Lawrence Berkeley National Laboratory, Berkeley, California 94305, USA}

\author{I. Belopolski}\affiliation {Joseph Henry Laboratory, Department of Physics, Princeton University, Princeton, New Jersey 08544, USA}

\author{N. Alidoust}\affiliation {Joseph Henry Laboratory, Department of Physics, Princeton University, Princeton, New Jersey 08544, USA}


\author{G. Bian}\affiliation {Joseph Henry Laboratory, Department of Physics, Princeton University, Princeton, New Jersey 08544, USA}


\author{P. P. Shibayev}\affiliation {Joseph Henry Laboratory, Department of Physics, Princeton University, Princeton, New Jersey 08544, USA}

\author{D. S. Sanchez}\affiliation {Joseph Henry Laboratory, Department of Physics, Princeton University, Princeton, New Jersey 08544, USA}

\author{Y. Ohtsubo}\affiliation {Synchrotron SOLEIL, Saint-Aubin-BP 48, F-91192 Gif sur Yvette, France}

\author{A. Taleb-Ibrahimi} \affiliation {Synchrotron SOLEIL, Saint-Aubin-BP 48, F-91192 Gif sur Yvette, France} \affiliation {UR1/CNRSSynchrotron SOLEIL, Saint-Aubin, F-91192 Gif sur Yvette, France}

\author{S. Basak}\affiliation {Department of Physics, Northeastern University, Boston, Massachusetts 02115, USA}

\author{W.-F. Tsai}\affiliation {Department of Physics, National Sun Yat-sen University, Kaohsiung 80424, Taiwan}

\author{H. Lin}
\affiliation{Centre for Advanced 2D Materials and Graphene Research Centre,
National University of Singapore, Singapore 117546}
\affiliation{Department of Physics, National University of Singapore,
Singapore 117542}

\author{Tomasz Durakiewicz}
\affiliation {Condensed Matter and Magnet Science Group, Los Alamos National Laboratory, Los Alamos, NM 87545, USA}


\author{R. J. Cava}\affiliation {Department of Chemistry, Princeton University, Princeton, New Jersey 08544, USA}

\author{A. Bansil}\affiliation {Department of Physics, Northeastern University, Boston, Massachusetts 02115, USA}


\author{F. C. Chou} \affiliation{Center for Condensed Matter Sciences, National Taiwan University, Taipei 10617, Taiwan}

\author{M. Z. Hasan}
\affiliation {Joseph Henry Laboratory, Department of Physics, Princeton University, Princeton, New Jersey 08544, USA}
\affiliation{Princeton Center for Complex Materials, Princeton University, Princeton, New Jersey 08544, USA}


\begin{abstract}
{
We report the evolution of the surface electronic structure and surface material properties of a topological crystalline insulator (TCI) Pb$_{1-x}$Sn$_x$Se as a function of various material parameters including composition $x$, temperature $T$ and crystal structure.
Our spectroscopic data demonstrate the electronic groundstate condition for the saddle point singularity, the tunability of surface chemical potential, and the surface states' response to circularly polarized light. 
Our results show that each material parameter can tune the system between trivial and topological phase in a distinct way unlike as seen in Bi$_2$Se$_3$ and related compounds, leading to a rich and unique topological phase diagram. Our systematic studies of the TCI Pb$_{1-x}$Sn$_x$Se are  valuable materials guide to realize new topological phenomena.}

\end{abstract}
\date{\today}
\maketitle

\section{INTRODUCTION}

A topological insulator (TI) material differs from a conventional insulator in the sense that a TI features metallic surface states that can only be removed by going through a topological quantum phase transition \cite{RMP, Zhang_RMP, Xia, David_nature, Chen_Science, SmB6_Hasan, Hasan QPT, Hasan_review}. Understanding the key parameters of a material that are relevant to its nontrivial topology is vital for utilizing the protected surface states in applications \cite{Xia, Hasan_review, David_nature, Chen_Science, graphene, graphene_th, graphene_Kondo, graphene_super, kim, Geim, RMP, Zhang_RMP, SmB6_Hasan, Hasan QPT, Cd3As2_Hasan}. A topological crystalline insulator (TCI) is a new symmetry protected topological phase that is protected by crystalline space group symmetries \cite{Liang NC SnTe, Liang PRL TCI, Fu_4}. The surface states of this novel topological phase are predicted to host many uniquely-new quantum phenomena, including surface spin filtering \cite{Fu_3}, strain-induced crystalline symmetry protected Chern currents \cite{Fu_1}, correlation physics due to surface electronic singularities \cite{Fu_2, stm}, none of which are possible in the much studied $\mathbb{Z}_2$ TI \cite{RMP, Zhang_RMP}. The realization of these proposals requires the ability to control a TCI material to be topological or non-topological as a function of various material parameters, and to understand the properties of the protected surface states under all these parameter conditions.

Recently, the TCI phase is experimentally realized in the Pb$_{1-x}$Sn$_x$Se(Te) material system  \cite{Ando, Suyang, PSS TCI, Valla, Ando_1}.
It is interesting to study the TCI materials at various sets of parameters (composition $x$, temperature $T$, lattice constant $a$, crystal structure, etc.).
Studying the Pb$_{1-x}$Sn$_x$Se(Te) system as a function of these material parameters is not just a material detail, but it is in fact important because varying  these parameters usually leads to a change of topology between a nontrival TCI and a trivial state. Furthermore, the exploration of TCI surface state properties and novel utilization are limited due to the lack of understanding of their material phase diagram.

In this report, we systematically study the surface electronic groundstate of the Pb$_{1-x}$Sn$_x$Se at various compositions $x$, temperatures $T$, and crystal structures by using angle resolved photoemission spectroscopy (ARPES), circular dichroism ARPES (CD-ARPES) and transport measurements. Our results reveal a remarkable tunablity in terms of the topological nature and the surface state electronic properties as a function of these parameters. Specifically, we show that by varying the composition $x$, Pb$_{1-x}$Sn$_x$Se undergoes an electronic band inversion at $x\sim0.20$ and a structural transition at $0.45\leq x\leq 0.75$, each of which alters the topology of the Pb$_{1-x}$Sn$_x$Se system in its unique way. Our studies show that the topological nature and the surface state electronic structure of Pb$_{1-x}$Sn$_x$Se are further distinctly sensitive to temperature. Moreover, we provide momentum-resolved evidence for the existence of a saddle point singularity in the mirror-protected surface states of Pb$_{1-x}$Sn$_x$Se.
These systematic studies of Pb$_{1-x}$Sn$_x$Se in terms of its topological nature and the surface state electronic structure as a function of the material parameters provide a valuable material reference to study various topological phenomena.

\section{METHODS}


\textbf{High-resolution ARPES measurements:}
Single crystals of Pb$_{1-x}$Sn$_{x}$Se used in these measurements were grown by standard growth method (see ref. \cite{SnSe_stru, SnSe_Stru_2}).
Low photon energy (15 eV-30 eV) and temperature dependent ARPES measurements for the low-lying electronic structures were performed at the Synchrotron Radiation Center (SRC) in Wisconsin and the Stanford Synchrotron Radiation Centre (SSRL) in California with R4000 electron analyzers whereas high photon energy ($\sim$60 eV) measurements were performed at the Beamlines 12 and 10 at the Advanced Light Source (ALS) in California, equipped with high efficiency VG-Scienta SES2002 and R4000 electron analyzers.  Samples were cleaved {\it in situ} and measured at 10-300 K in vacuum better than $1\times10^{-10}$ torr. They were found to be very stable and without degradation for the typical measurement period of 20 hours. Energy and momentum resolution were better than 15 meV and 1\% of the surface Brillouin zone (BZ), respectively.

\textbf{Surface deposition:}
Sn deposition ARPES measurements were done in The CASSIOPEE beamline, Soleil, France from Sn source, which was thoroughly degassed before the experiment. Pressure in the experiment chamber stayed $1\times10^{-10}$ torr. The deposition rate (\AA/min) was monitored using  a commercial quartz thickness monitor (Leybold In con Inc., Model XTM/2). The deposition rate of Sn was 0.3 \AA/min.

 \textbf{CD-ARPES measurements:}
CD-ARPES measurements for the low energy electronic structure were performed at the Synchrotron Radiation Center (SRC), Wisconsin, equipped with high efficiency VG-Scienta SES2002 electron analyzer.
The polarization purity is better than 99\% for horizontal polarization (HP) and better than 80\% for RCP and LCP.
Samples were cleaved {\it in situ} and measured at 20 K in a vacuum better than 1 $\times$ 10$^{-10}$ torr.
Energy and momentum resolution were better than 15 meV and 1\% of the surface Brillouin zone (BZ), respectively.

\textbf{Synchrotron-based XRD:}
Synchrotron-based X-ray diffraction (XRD) measurements were performed with X-ray beamline at the Spring-8 light source which has access to hard X-rays. The synchrotron X-ray diffraction patterns were analyzed using the General Structure Analysis System (GSAS) program \cite{SXRD} following the Rietveld profile refining method. The final refinements were carried out assuming a cubic symmetry with a space group of \textit{Fm-3m} and taking the pseudo-Voigt function for the peak profiles. 

\textbf{First-principles calculations:}
The first-principles calculations are carried out within the framework of the density functional theory (DFT) 
using projector augmented wave method \cite{paw} as implemented in the VASP package \cite{vasp}. The generalized gradient approximation (GGA) \cite{gga} is used to model exchange-correlation effects. The spin orbital coupling (SOC) is included in the self-consistent cycles. The cutoff energy 260 eV is used for both PbSe and SnSe systems. 
To capture the physical essence of the surface state electronic structure of the TCI phase in an alloy Pb$_{1-x}$Sn$_{x}$Se system, we take SnSe and assume that SnSe crystalizes in the FCC structure and then calculate the $(001)$ surface electronic structure. We also calculate the electronic structure of PbSe (FCC) and SnSe (orthorhombic) using experimentally determined crystal structure and lattice constant.  For all calculations, the $(001)$ surface is modeled by periodically repeated slabs of 48-atomic-layer thickness with  24 $\textrm{\AA}$ vacuum regions and use a 12 $\times$ 12 $\times$ 1 Monkhorst-Pack  $k-$point mesh over the (BZ). 

\textbf{$k\cdot p$ model:}
We use an effective surface $k\cdot p$ Hamiltonian \cite{kdotp} of topological crystalline insulator on $(001)$ plane to obtain the surface bands dispersion. 

\begin{figure*}
\centering
\includegraphics[width=17cm]{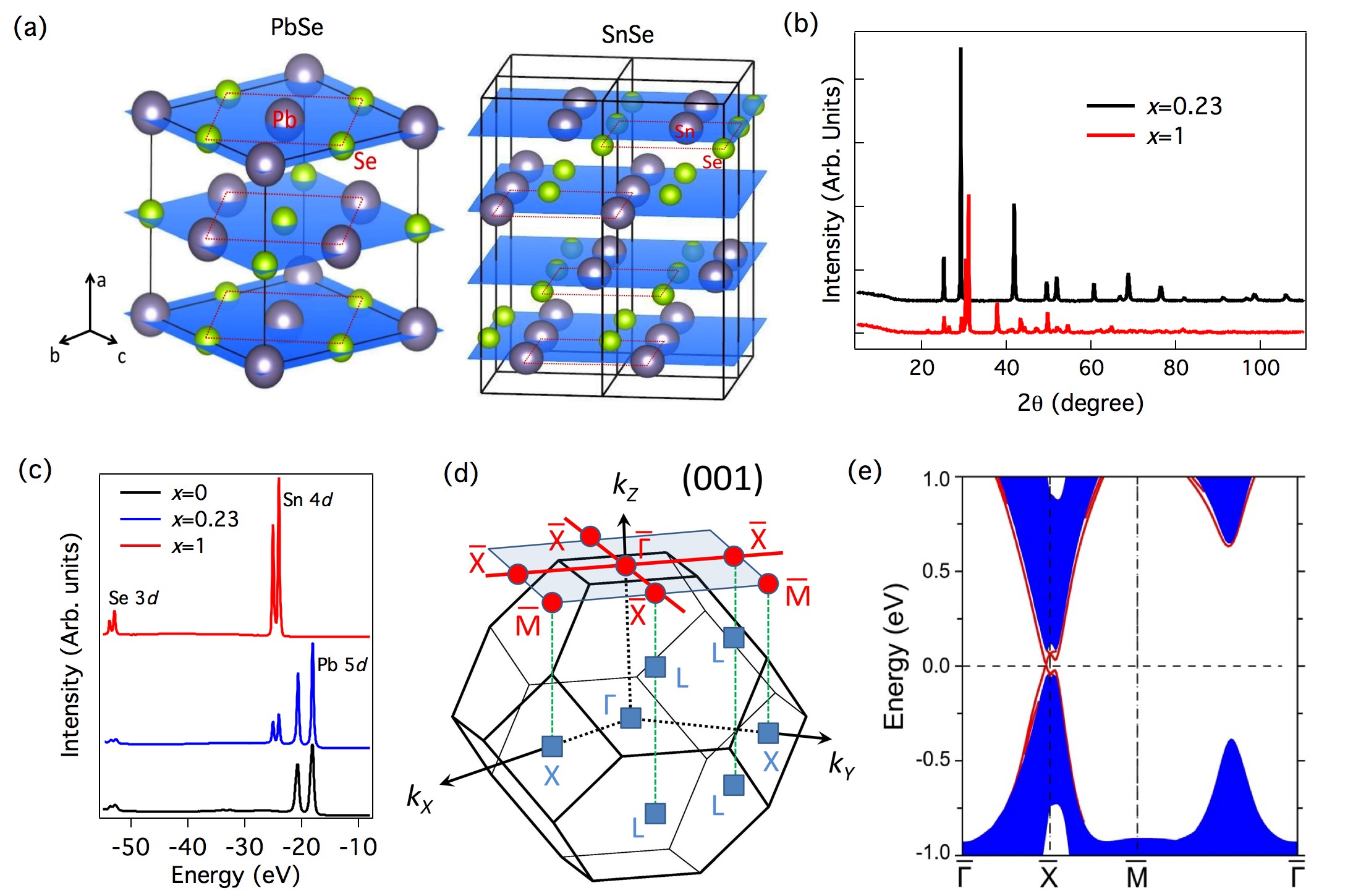}
\caption{\label{phase_fig}
(a)  Crystal structure of PbSe with ideal NaCl type and SnSe with orthorhombic distortion. 
(b) X-ray power diffraction patterns for Pb$_{0.77}$Sn$_{0.23}$Se and SnSe systems. (c) Core level spectroscopic measurements of PbSe, Pb$_{0.77}$Sn$_{0.23}$Se and SnSe systems. Various core level energy peaks are marked on the curves. (d) The first Brillouin of Pb$_{1-x}$Sn$_{x}$Se. The mirror planes are projected onto the (001) crystal surface as the $\bar{{X}}$-$\bar\Gamma$-$\bar{{X}}$ momentum cut mirror lines (shown by red solid lines). (e) First-principles calculations of band dispersion along high-symmetry momentum space cuts. Red lines and blue areas represent the surface and bulk bands, respectively.}
\end{figure*}

\begin{SCfigure*}
\centering
\includegraphics[width=13.5cm]{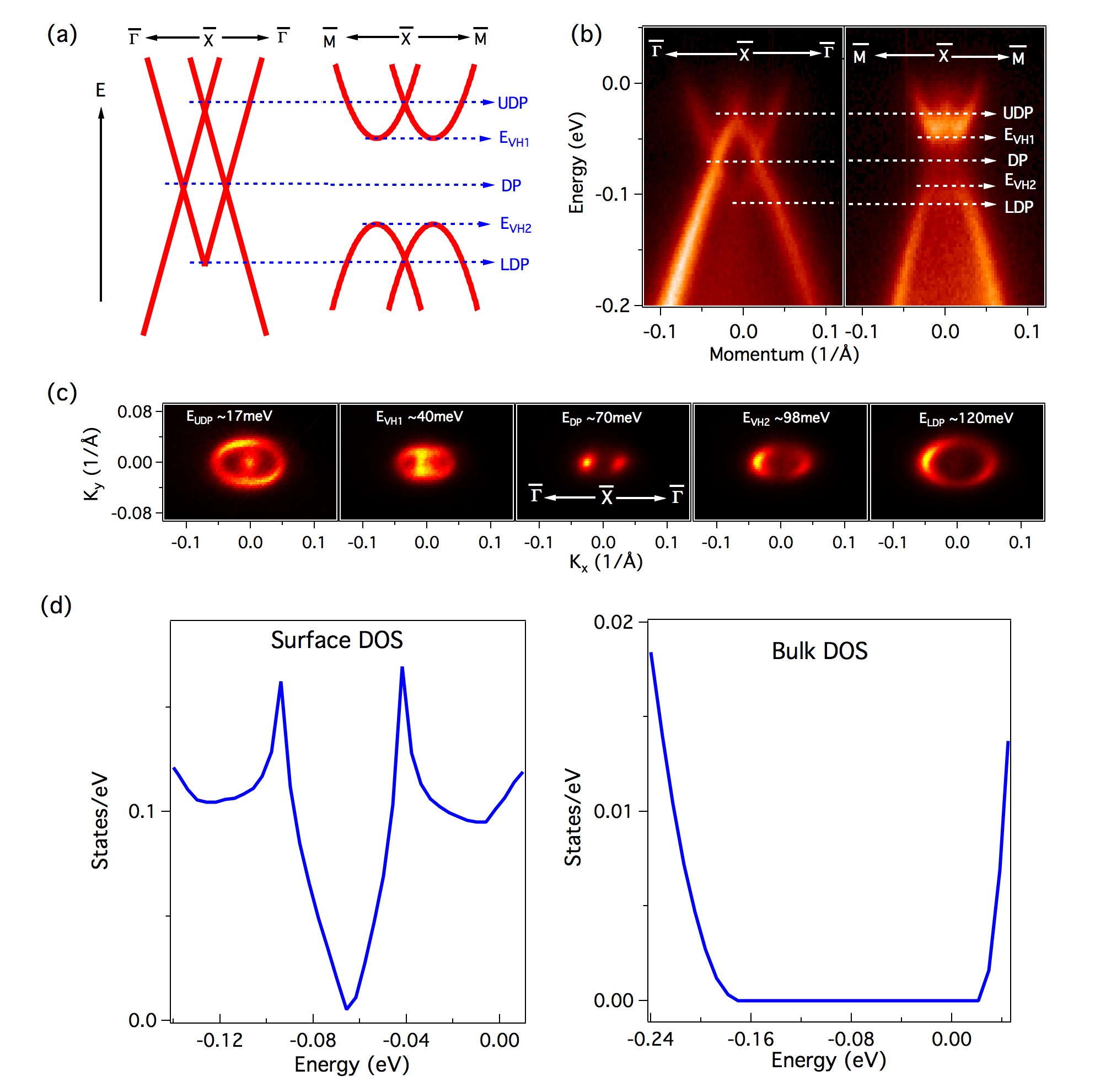}
\caption{\label{saddle}{Observation of the saddle point singularity.} 
(a) Schematics of surface band dispersion of the TCI phase along the mirror line $\bar\Gamma$-$\bar{{X}}$-$\bar\Gamma$ and the perpendicular to the mirror line $\bar{{M}}$-$\bar{{X}}$-$\bar{{M}}$ momentum space cut-directions. Five important features of the surface state, including Dirac point of the upper Dirac cones (UDP), van Hove singularity of the upper Dirac cones (VH1), two Dirac points along the $\bar\Gamma$-$\bar{{X}}$-$\bar\Gamma$ mirror line (DP), van Hove singularity of the lower Dirac cones (VH2) and Dirac point of the lower Dirac cones (LDP) are marked. (b) ARPES measured dispersions along the $\bar\Gamma$-$\bar{{X}}$-$\bar\Gamma$ and $\bar{{M}}$-$\bar{{X}}$-$\bar{{M}}$ momentum space cut-directions, where five features are also marked. (c) Experimental observation of the Lifshitz transition - the binding energies are noted on the constant energy contours.  (d) Calculated density of state (DOS) for the surface states and the bulk bands using the $k\cdot p$ model \cite{kdotp}.  ARPES data presented here were measured at low temperature (T $\sim$ 10 K) with photon energy of 18 eV at SSRL BL 5-4.}
\end{SCfigure*}

\section{RESULTS AND DISCUSSION}

\subsection{Crystal structure and Brillouin zone symmetry}

Since the nontrivial topology in a TCI state fundamentally relies on crystalline symmetries, we first characterize the crystal structure of Pb$_{1-x}$Sn$_x$Se. Fig. \ref{phase_fig}a shows the crystal structure of PbSe ($x=0.0$) and SnSe ($x=1$). 
Based on X-ray diffraction (XRD) data (Fig. 1b), the $x=0.0$ sample has a face centered cubic rocksalt structure (Fm$\bar{3}$m, \#225). The unit cell is shown by the red square and includes one Pb/Sn site in the center and one Se site shared by the corners, see Fig. \ref{phase_fig}a left. The $x=1$ sample has a primitive orthorhombic structure (Pnma, \#62) based on the XRD (Fig. 1b). In this structure, none of the lattice vectors $a$, $b$ or $c$ are equal and the Pb/Sn site is displaced from the center of the unit cell, see Fig. \ref{phase_fig}a right.

We study core level spectra in a binding energy range 10 eV - 56 eV, Fig. \ref{phase_fig}c. For the $x = 0$ composition, we observe the Pb 5$d$ ($\sim$20eV) and Se 3$d$ ($\sim$55eV) core level peaks. For $x=0.23$, we observe the Pb 5$d$ ($\sim$20eV), Sn 4$d$ ($\sim$24eV) and Se 3$d$ ($\sim$55eV) core level peaks. For $x=1$, we observe the Sn 4$d$ ($\sim$24eV) and Se 3$d$ ($\sim$55eV) states. The sharp core level peaks indicate the high quality of the samples used for our measurements .
For $x$ values where the crystal has rocksalt structure, the electronic structure is known to have a direct band gap at the $L$ points \cite{SnSe_stru}. These are projected to the $\bar{X}$ points on the (001) surface, see Fig. 1d. As $x$ varies within the rocksalt range, the system undergoes a band inversion at each $L$ point, see Fig. 1e. The inverted Pb$_{1-x}$Sn$_x$Se(Te) compositions are expected to show topological surface states protected by mirror symmetries. These are the signatures of the TCI state \cite{Liang NC SnTe}.

\begin{figure*}
\centering
\includegraphics[width=18.0cm]{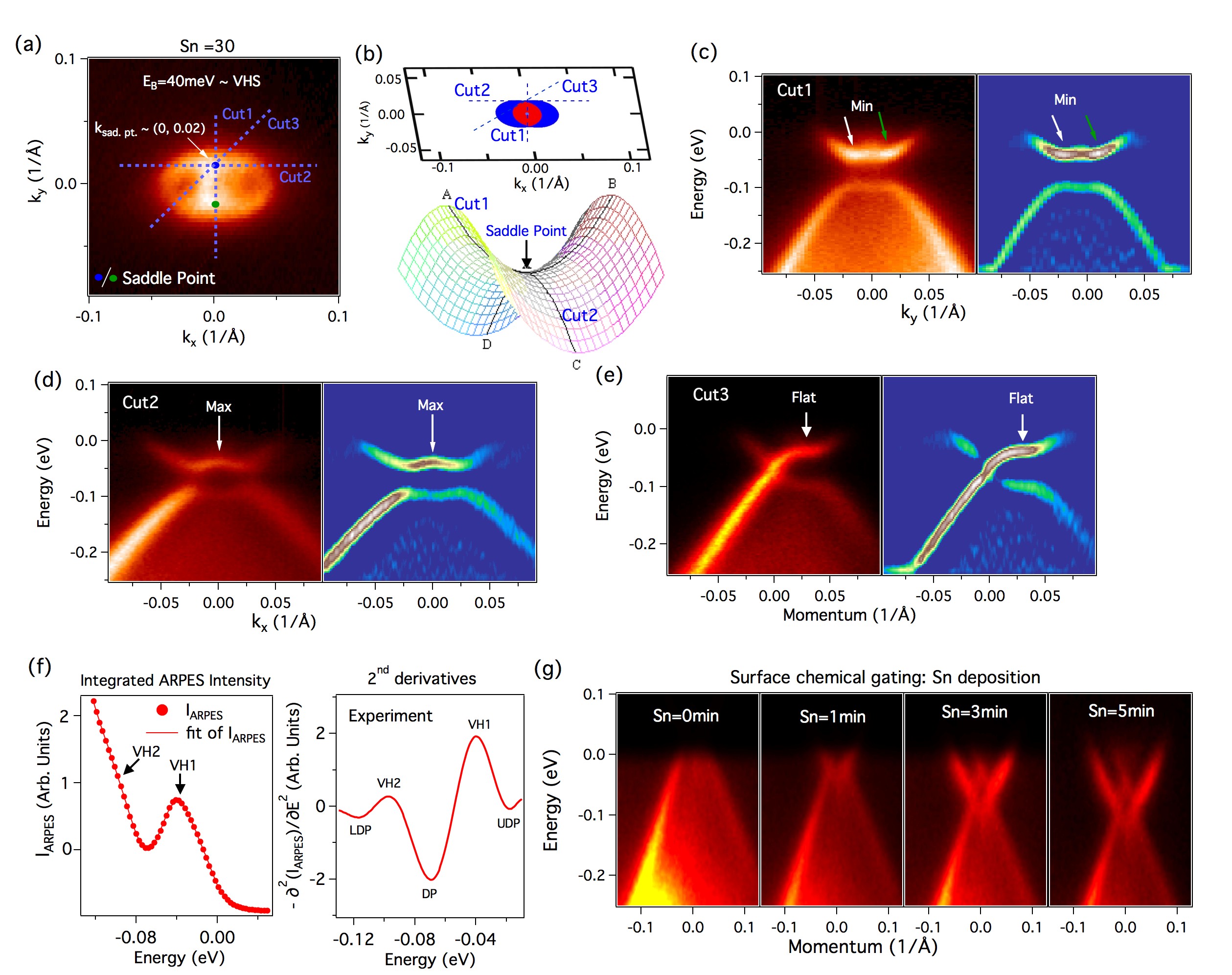}
\caption{\label{saddle_1}{Observation of the saddle point curvature.} 
(a) ARPES constant energy contour in the vicinity of an $\bar{{X}}$ point in the (001) surface BZ of Pb$_{0.70}$Sn$_{0.30}$Se at binding energy 40 meV, which corresponds to the saddle point energy of the upper Dirac cones. The $\bar{{X}}$ points locate at $(k_x,k_y)=(\pm0.7$ $\textrm{\AA}^{-1},0)$ or $(0, \pm0.7$ $\textrm{\AA}^{-1})$. 
The momentum space coordinate of one $\bar{{X}}$ point is redefined to be $(k_x,k_y)=(0,0)$ for simplicity of presentation. 
The blue and green dots denote the momentum space locations of the two surface saddle points. The blue dotted lines indicate the momentum space cut-directions for cuts1 (along the $\bar{{M}}$-$\bar{{X}}$-$\bar{{M}}$ momentum space cut), cut2 (parallel to the $\bar\Gamma$-$\bar{{X}}$-$\bar\Gamma$ momentum space cut), and cut3 (in between cut1 and cut2), which are centered at the blue dot. (b) Calculated surface state constant energy contour at the saddle point singularity energy (top), where outer blue elliptical and inner red circular features represent the contour from outer cone and inner child cone respectively, and a three-dimensional schematic of a saddle point (bottom). (c-e) ARPES dispersion maps (left) and their second derivative images along cuts 1, 2, and 3. The white and green arrows point the saddle points (blue and green dots in Panel (a). (f) Momentum ($k_x$ and $k_y$) integrated ARPES intensity as a function of binding energy (left). 2$^{nd}$ derivative of the ARPES intensity with respect to binding energy is presented to further highlight the features (right panel). The upper Dirac point (UDP), upper van Hove singularity (VH1), Dirac point (DP), lower van Hove singularity (VH2) and lower Dirac point (LDP) are marked. (g) ARPES dispersion maps upon \textit{in situ} Sn deposition on the Pb$_{0.70}$Sn$_{0.30}$Se surface. The dosage (time) for Sn deposition is noted. A different batch of sample, which is $p-$type with the chemical potential located below the Dirac points, is used for the Sn deposition data shown in this panel. ARPES data presented here were measured at low temperature (T $\sim$ 10 K) with photon energy of 18 eV.}
\end{figure*}


\subsection{Tunable topological surface state and saddle point singularity.}
We systematically study the electronic structure of the symmetry-protected topological surface states in Pb$_{0.70}$Sn$_{0.30}$Se ($x=0.3$). Previously, although topological surface states have been observed by ARPES in Pb$_{0.67}$Sn$_{0.23}$Se ($x=0.23$), the main purpose was to demonstrate the TCI state \cite{PSS TCI}. However, the saddle point singularity, the surface chemical potential tunability and the spin-orbit polarization have \textit{not} been experimentally characterized with momentum resolution. Here, we focus on these key properties of the symmetry-protected topological surface states, which are crucial for the utilization of the TCI phase in Pb$_{1-x}$Sn$_x$Se. 

In Figs.~\ref{saddle}a, b we show the dispersion of the topological surface states in Pb$_{0.70}$Sn$_{0.30}$Se along the mirror line, $\bar{\Gamma}$-$\bar{X}$-$\bar{\Gamma}$, as well as along the $\bar{M}$-$\bar{X}$-$\bar{M}$ direction, which is perpendicular to the mirror line. We observe two surface Dirac cones along the mirror line, on opposite sides of the $\bar{X}$ point. We find that these two Dirac cones are near each other in momentum space, so they hybridize with each other at binding energies far from the energy of the Dirac point, $E_{\textrm{DP}}$. In Fig.~\ref{saddle}c we show constant energy contour maps at different binding energies, $E_{\textrm{B}}$. We find that the Dirac points are at $E_{\textrm{B}}=E_{\textrm{DP}}=70$ meV. At energies $E_{\textrm{B}}$ slightly above or below $E_{\textrm{DP}}$, the constant energy contour evolves from two points into two unconnected pockets. As we scan in $E_{\textrm{B}}$, we find that the size of the pockets increases until they come together and undergo a Liftshitz transition. A Lifshitz transition in the band structure is associated with a van Hove singularity in the density of states (DOS), since the band structure is nearly flat around the saddle point where the bands touch \cite{Fu_4, Lifshitz}. This observation is also predicted by our first-principles calculation of the surface and bulk DOS, see Fig.~\ref{saddle}d. For still higher $E_{\textrm{B}}$, the constant energy contour changes into two concentric contours both enclosing the $\bar{X}$ point. The inner contour shrinks with increasing $E_{\textrm{B}}$ until it forms an upper Dirac point at large $E_{\textrm{B}}$. In summary, there are five important energies in the surface state band structure: the Dirac point (DP), the Lifshitz transition energy for the upper Dirac surface states ($E_{\textrm{VH1}}$) and lower Dirac surface states ($E_{\textrm{VH2}}$), as well as the Dirac points for the upper and lower Dirac surface states (UDP and LDP), all of which are noted in Figs.~\ref{saddle}a-c.

\begin{figure}
\centering
\includegraphics[width=8.800cm]{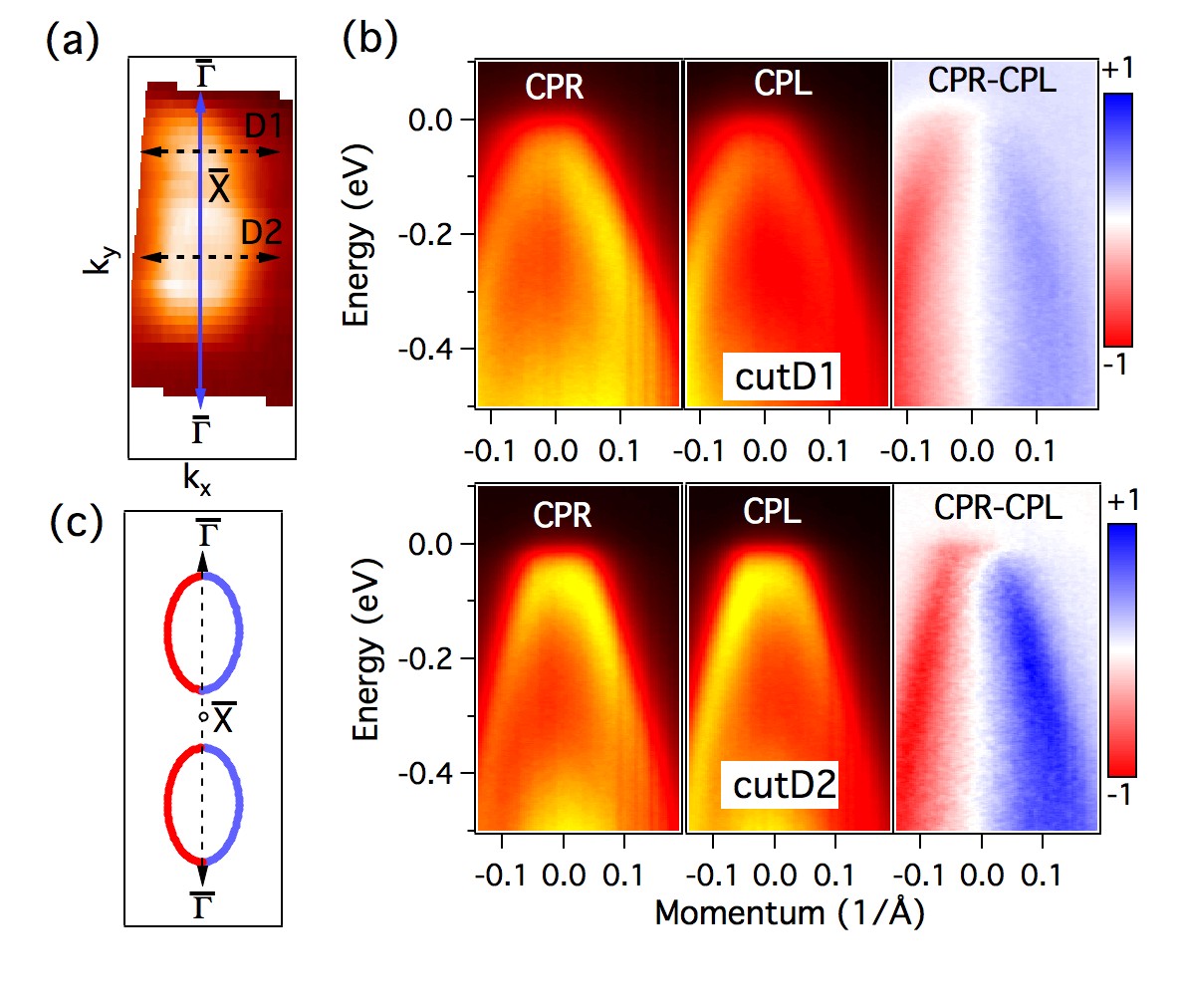}
\caption{\label{CD}
(a) Fermi surface plot of Pb$_{0.70}$Sn$_{0.30}$Se. (b) Circular dichroism measurements of Pb$_{0.70}$Sn$_{0.30}$Se. The ARPES spectra are measured using the light of circularly polarized right (CPR) and left (CPL) along two different momenta marked as D1 and D2 in (a). The circular dichroism is the difference between those two spectra (CD= CPR$-$CPL). These data were measured at temperature of 20 K with photon energy of 18 eV at SRC U9 VLS-PGM beamline. (c) Schematic view of the CD results.}
\end{figure}

\begin{SCfigure*}
\centering
\includegraphics[width=13.50cm]{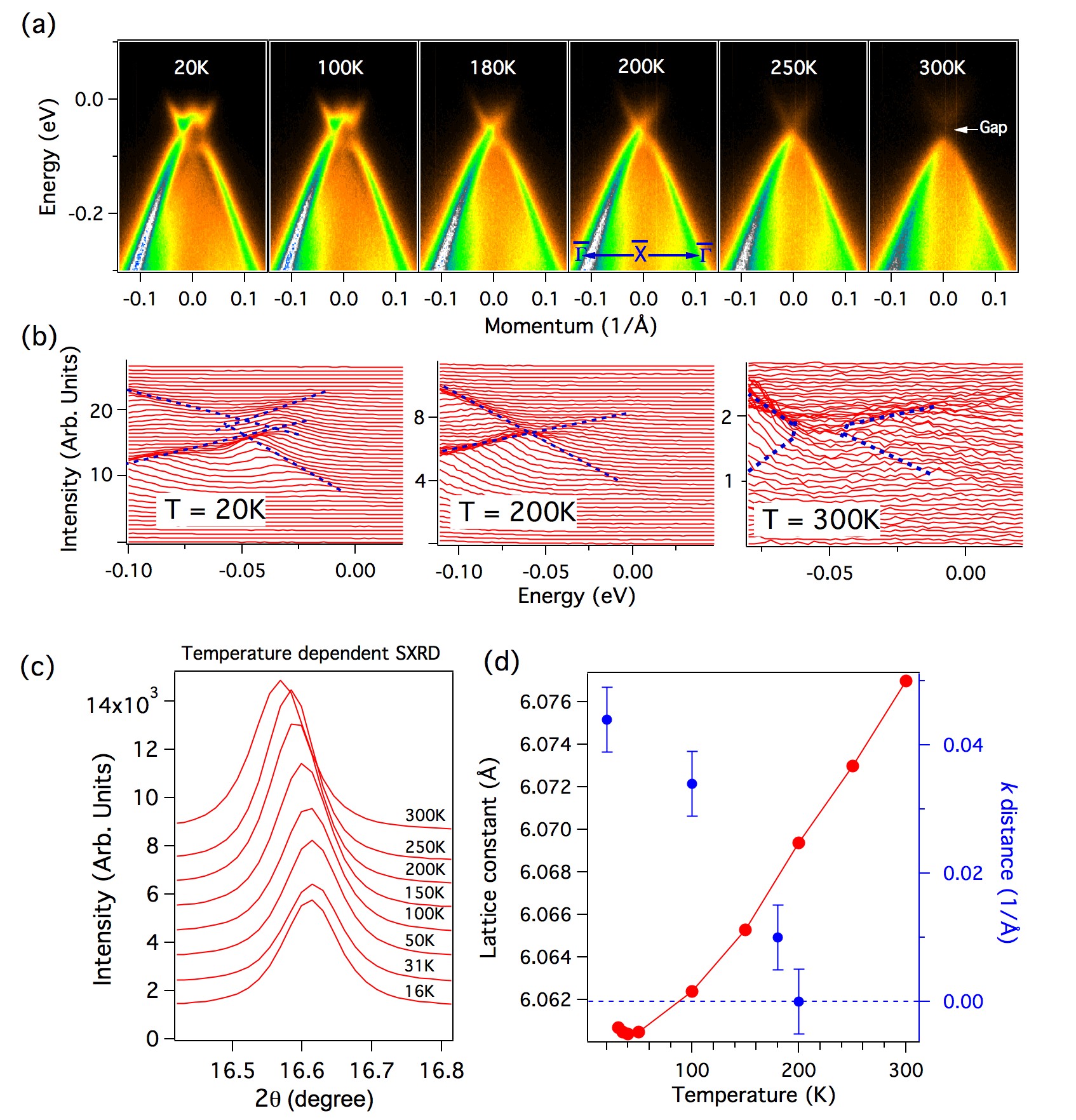}
\caption{\label{temp}
Thermal evolution of the topological surface states (TCI $\rightarrow$ Trivial).
 (a) Dispersion maps along the mirror line $\bar\Gamma$-$\bar{{X}}$-$\bar\Gamma$ at different temperatures. 
These data are measured with photon energy of 18 eV at SSRL BL 5-4. (b) Energy dispersion curves (EDCs) of the selected spectra shown in (a). The measured temperatures are noted on the plots. The blue dash lines are guide to the eye. 
 (c) Synchrotron-based temperature dependent X-ray diffraction (SXRD) measurements for Pb$_{0.70}$Sn$_{0.30}$Se. The peak is observed to shifts towards the lower angles with increasing temperature, which confirms the picture of the thermal expansion of lattice. (d) Experimentally measured values of the lattice constant at different temperatures (left axis). The momentum distance between two Dirac points as a function of temperature is plotted in the right axis. The error bar represents an uncertainty of estimating the momentum distance between two Dirac points.} 
\end{SCfigure*}

\begin{figure*}
\centering
\includegraphics[width=17.2cm]{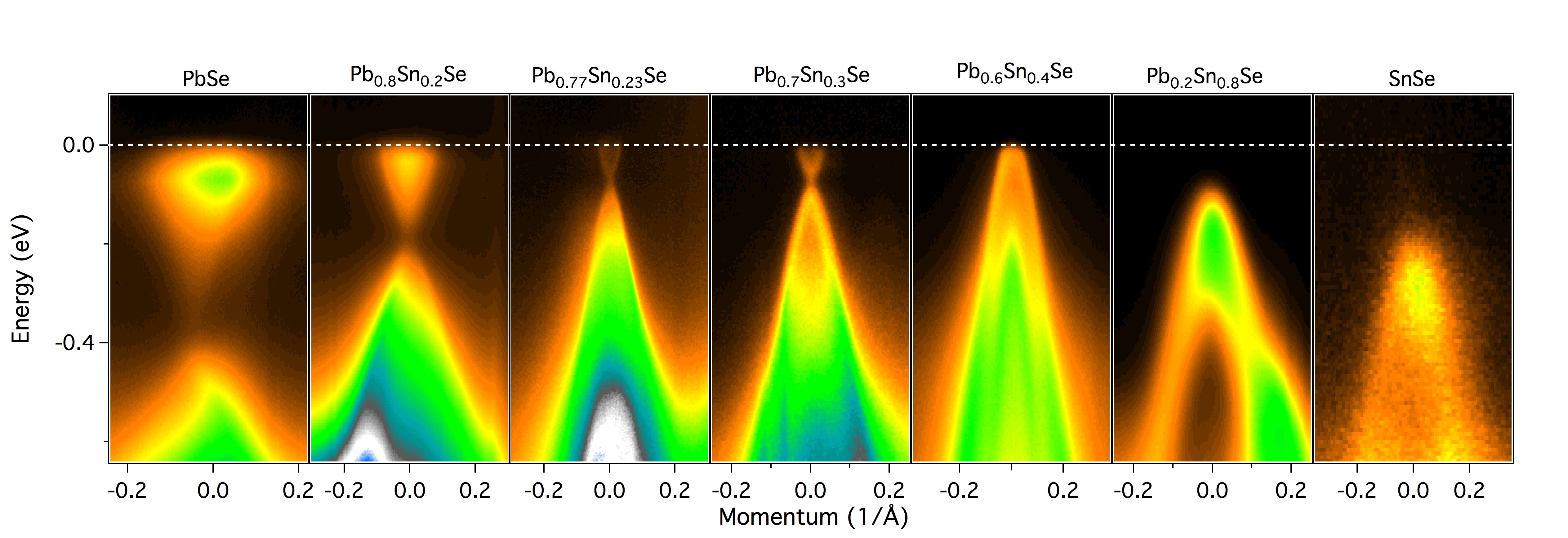}
\caption{\label{ARPES}{Topological phase transition in Pb$_{1-x}$Sn$_{x}$Se.}  
 ARPES measured dispersion map of PbSe, Pb$_{0.8}$Sn$_{0.2}$Se, Pb$_{0.77}$Sn$_{0.23}$Se, Pb$_{0.7}$Sn$_{0.3}$Se, Pb$_{0.6}$Sn$_{0.4}$Se, Pb$_{0.3}$Sn$_{0.8}$Se and SnSe. Note that PbSe, Pb$_{0.2}$Sn$_{0.8}$Se and SnSe are observed to be insulators.
The Dirac-like surface state dispersion is observed in Pb$_{0.77}$Sn$_{0.23}$Se, Pb$_{0.7}$Sn$_{0.3}$Se, Pb$_{0.6}$Sn$_{0.4}$Se while no surface state is observed in Pb$_{0.8}$Sn$_{0.2}$Se. 
 PbSe, Pb$_{0.8}$Sn$_{0.2}$Se, Pb$_{0.77}$Sn$_{0.23}$Se, Pb$_{0.7}$Sn$_{0.3}$Se and Pb$_{0.6}$Sn$_{0.4}$Se are measured at low temperature ($\sim$ 20K). The chemical composition of the measured compounds are noted on each panels. Note that Pb$_{0.6}$Sn$_{0.4}$Se sample is $p-$type where the Fermi level is below the Dirac point.
Energy gap is observed in Pb$_{0.2}$Sn$_{0.8}$Se and SnSe measured at temperature of 100K. All these spectra were measured parallel to the $\bar{{M}}$-$\bar{{X}}$-$\bar{{M}}$ momentum cut.}
\end{figure*}

In order to identify these key features, especially the interesting van Hove singularities, we further study the surface states' electronic structure. We focus on the constant energy contour at the Lifshitz transition energy, $E_{\textrm{B}}=40$ meV (Fig. \ref{saddle_1}a). The two dots (green and blue) in Fig. \ref{saddle_1}a mark the momentum space locations where the two unconnected contours merge. Their energy and momentum space coordinates are experimentally identified to be $(E_{\textrm{B}}, k_x, k_y)=(40$ meV$, 0, \pm0.02$ $\textrm{\AA}^{-1})$. To experimentally establish the saddle point band structure, we focus on the blue dot in Fig. \ref{saddle_1}a and study the energy-momentum dispersion cuts along three important momentum space cut-directions, labelled as cut 1, 2, and 3. Cut 1 and 2 (Figs. \ref{saddle_1}c and d) show the dispersion along the horizontal ($k_x$) and vertical ($k_y$) directions across the blue dot. Interestingly, the blue dot is found to be a local band structure minimum along cut 1 shown in Fig. \ref{saddle_1}c, whereas it is a local maximum along cut 2 (Fig. \ref{saddle_1}d). Observation of a local minimum and local maximum at the same momentum space location (the blue dot) unambiguously  shows that it is a surface band structure saddle point. The observation of a surface momentum-space saddle point implies that there exists certain intermediate cut-directions (between cut 1 and 2), where the surface band structure is flat in the vicinity of the blue dot. Indeed, as shown in Fig. \ref{saddle_1}e, for cut 3, we found that the surface states are nearly flat near the saddle point. The observed flat band structure (along cut 3) is associated with a divergence of the surface density of states (DOS), a surface van Hove singularity (VHS). Indeed, a pronounced peak is observed at the energy corresponding to the saddle point, namely $E_{\textrm{B}}=40$ meV, as labelled by ``VH1'' in Fig. \ref{saddle_1}f. Additionally, we observe a significant dip of angle-integrated photoemission intensity at the binding energy of $E_{\textrm{B}}=70$ meV, which corresponds to the energy of the Dirac points (DP). In order to better highlight the other three features (UDP, LDP, VH2), we present a second derivative of the angle-integrated photoemission intensity (Fig. \ref{saddle_1}f right), where all five features are identified and show qualitative correspondence with the theoretical calculated results in Fig.~\ref{saddle}d. 


Here, we demonstrate the ability to control the surface chemical potential via surface chemical deposition. As shown in Fig. \ref{saddle_1}g, Sn atoms are deposited onto the surface of a $p$-type Pb$_{0.70}$Sn$_{0.30}$Se sample, whose native chemical potential is found to be below the Dirac point. The deposition rate of Sn on the surface of the sample is estimated to be about 0.3 \AA \hspace{0.1cm} per minute. Our Sn surface deposition shows that the surface chemical potential can be shifted across the energies of the Dirac point and the van Hove singularity. 

We further perform circular dichroism (CD) ARPES measurements in attempt to probe  the spin-orbit polarization of the topological surface states. Previously, spin-resolved ARPES results were reported on TCI states \cite{Suyang, PSS_TCI_spin}. CD-ARPES not only brings insight to the spin-orbit polarization of a topological surface state, as was already shown in TI Bi$_2$Se$_3$ \cite{Gedik, Park}, but also holds promise to manipulate the spin texture via circular polarized photon excitations \cite{laserspinarpes, Neupane_CD, Rader}. However, there have been no CD-ARPES measurements performed on the TCI Pb$_{1-x}$Sn$_x$Se(Te) systems. Fig. \ref{CD}b shows ARPES dispersion maps of surface states measured using right circularly polarized (RCP) light and left circularly polarized (LCP) light for a $p$-type Pb$_{0.7}$Sn$_{0.3}$Se. The directions of the dispersion cuts are noted as D1 and D2 in Fig. \ref{CD}a. A clear surface state CD response on the photoelectron signal from the lower Dirac cone surface states is observed where the $+k$ Dirac branch shows stronger response for RCP light and the $-$k Dirac branch shows  stronger response for LCP light (Fig. \ref{CD}b) for both lower Dirac cones. The magnitude of the CD response signal defined, as $I_{CD}$=($I_{RCP}-I_{LCP}$)/($I_{RCP}+I_{LCP}$) is observed to be about 20\% for incident photons with energy 20 eV at binding energy 100 meV, well below the chemical potential. The CD signal observed in Fig. \ref{CD}b is consistent with the expected spin-orbital texture of the surface states (shown schematically in Fig. \ref{CD}c). 

\subsection{Topological phase transitions and topological phase diagram for Pb$_{1-x}$Sn$_x$Se.}


We now study the topological nature of Pb$_{1-x}$Sn$_x$Se as a function of the composition ($x$), temperature ($T$), and crystal structure. We note that the temperature and composition dependent studies of Pb$_{1-x}$Sn$_x$Se (with 0$\leq x \leq$ 0.37) have been discussed in Ref. \cite{Wojek_1}.
Our goal here is to map out an entire topological phase diagram for Pb$_{1-x}$Sn$_x$Se (0$\leq x \leq$ 1.0), and to study how the topological surface states arise and disappear as the system goes through various topological phase transitions. 
In Fig. \ref{temp}a and b, we present ARPES spectra of Pb$_{0.70}$Sn$_{0.30}$Se at different temperatures. At a low temperature, $T=20$ K, we observe two Dirac cones near the $\bar{X}$ point along the $\bar{\Gamma}$-$\bar{X}$-$\bar{\Gamma}$ cut, which are associated with the nontrivial TCI phase. We repeat this measurement at 100 K, 180 K and 200 K, and we find that as the temperature rises the two Dirac points move closer to each other. At $\sim 250$ K, we see that the two cones merge at the $\bar{X}$ point. The momentum space distance between the two Dirac points are measured to be $0.05$ $\textrm{\AA}^{-1}$, $0.03$ $\textrm{\AA}^{-1}$, $0.01$ $\textrm{\AA}^{-1}$, and $0$ $\textrm{\AA}^{-1}$ at 20 K, 100 K, 180 K and 200 K, respectively. Finally, at 300 K we observe that a gap opens in the surface states, reflecting the transition into a topologically trivial phase (rightmost panel, Fig. \ref{temp}a-b).


 To further study the thermodynamic evolution of the surface state electronic structure, we perform synchrotron-based X-ray diffraction measurements on the sample as a function of temperature. As shown in Fig. \ref{temp}c, the peak of the diffraction $2\theta$ angle is clearly found to move towards smaller values with rising temperature showing that the lattice constants increase with increasing temperature (Fig. \ref{temp}d). The variation of the momentum distance between two Dirac points with temperature (Fig.
\ref{temp}d, right  axis) shows an evolution of the trivial state from TCI. This is qualitatively consistent with thermodynamic expansion of the lattice constant. 



\begin{figure*}
\centering
\includegraphics[width=16cm]{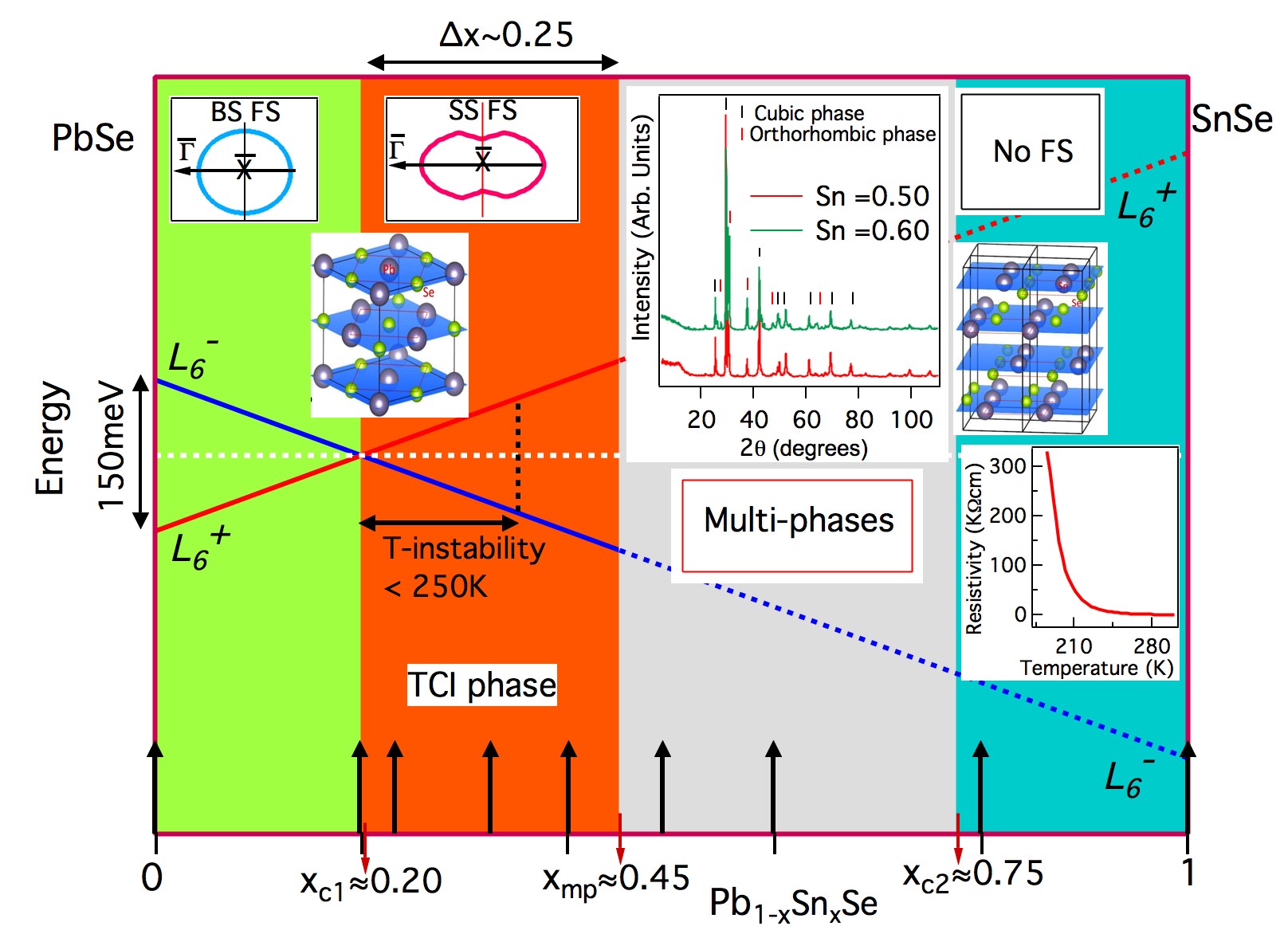}
\caption{\label{phase}
Topological phase diagram of the Pb$_{1-x}$Sn$_{x}$Se system. For composition range of $0<x<0.45$, the system is in the single crystalline FCC phase. The bulk band of Pb$_{1-x}$Sn$_{x}$Se undergoes a band inversion with Pb/Sn substitution. 
The critical composition $x_{c1}$ is $\sim0.20$ depending on the temperature. The conduction and valence band states representing odd and even parity eigenvalues are marked as 
$L_6^-$ and $L_6^+$, respectively. For composition range of $0.45<x<0.75$, the system shows multi-structural-phase (cubic and orthorhombic phases coexist. See the XRD data in the inset for $x=0.5$ and $0.6$). The upper insets are schematic Fermi surface plots around the $\bar{{X}}$ point. The inset in the bottom right conner shows the resistivity measurements on SnSe, which proves its insulator nature. BS FS and SS FS denote the bulk state Fermi surface and surface state Fermi surface, respectively. The arrows at bottom note the compositions where our ARPES studies have been performed. For composition range of $0.75<x<1$, the system is in a single crystalline orthorhombic phase.}

\end{figure*}

We study the TCI phase in Pb$_{1-x}$Sn$_x$Se as a function of composition $x$. 
Fig. \ref{ARPES} shows ARPES measurements of the low energy states of various compositions together with the two end compounds, namely PbSe and SnSe. For PbSe, we observe low-lying bulk conduction and valence bands with a clear band-gap of $\sim0.15$ eV. This suggests that the system is topologically trivial for $x=0$. As $x$ is increased (see Fig. \ref{ARPES} ), the low-lying bulk bands are observed to approach each other, and eventually their energy levels invert. We find that the critical composition for band inversion is at $x \sim0.20$. 
It is interesting to note that the spectrum of the $x = 0.2$ composition at low temperature ($\sim$ 20 K) shows the presence of weak spectral weight in the bulk band gap and at the boundary of the bulk valence and conduction bands. These may be preformed surface states in the vicinity of topological phase transition. Analogous preformed states have recently been observed experimentally \cite{Suyang_preformed, Vidya}.
Next, we consider the other end compound, SnSe, as shown in the right panel in Fig. \ref{ARPES}. At the Fermi level, no electronic states are observed. Instead, a fully gapped electronic structure with the chemical potential inside the band gap is found. This is because SnSe is in primitive orthorhmobic structure \cite{SnSe Crystal, SnSe_distor}. This crystal structure breaks both inversion and mirror symmetry, removing the fundamental symmetries which support the TCI phase. Our observation of an insulating state in SnSe is consistent with previous studies where the band gap is reported to be as large as $\sim1$ eV \cite{SnSe_gap}. Our systematic ARPES studies and transport measurements show a rich topological phase diagram in Pb$_{1-x}$Sn$_x$Se. A summary is presented in Fig. \ref{phase}. The blue and red lines represent the energy levels of the lowest lying bulk conduction and valence bands. Starting from PbSe ($x=0$), the system has a non-inverted band-gap of $\sim0.15$ eV. As $x$ increases, band inversion takes place and the system enters the TCI phase. The inverted band-gap increases until the system enters the multi-(crystal structure)-phase regime at $x\gtrsim0.45$, where cubic and orthorhombic structures coexist. Finally, for $x\gtrsim0.75$, the system becomes a large band gap trivial insulator with primitive orthorhombic structure. Distinct phase transitions are observed, labelled $x_{\textrm{c1}}$, $x_{\textrm{mp}}$ (where `mp' stands for multi-phase) and $x_{\textrm{c2}}$. The first transition ($x_{\textrm{c1}}\gtrsim0.2$) is due a decrease in the lattice constant, which increases the effective spin-orbit strength, while the second transition ($x_{\textrm{mp}}\gtrsim0.45$) corresponds to the coexistence of  cubic and orthorhombic structures and the third transition ($x_{\textrm{c2}}\gtrsim0.75$) is the result of a drastic change in crystal structure to orthorhombic phase. 
Therefore, our experimental data reveal a delicate relationship among lattice constant, band gap, effective spin-orbit coupling strength and crystal structure associated with the topological phase transition in Pb$_{1-x}$Sn$_{x}$Se.

\bigskip

\section{CONCLUSION}
In conclusion, by utilizing ARPES, CD-ARPES and transport measurements, we provide a comprehensive topological phase diagram for Pb$_{1-x}$Sn$_x$Se as a function of composition ($x$), temperature ($T$) and crystal structure. We show that each of these three material parameters can tune the system between a trivial and topological phase differently, giving rise to a rich topological phase diagram. 
Moreover, we reveal the momentum space nature of interconnectivity of the Fermi surface pockets leading to a saddle point singularity within the topological surface state in the TCI. 
We further show that the measured momentum-integrated density of states exhibits pronounced peaks at the saddle point energies, demonstrating the van Hove singularities (VHSs) in the topological surface states, whose surface chemical potential, as we have shown, can be tuned via surface chemical gating.
Our systematic measurements represent a material guide for TCIs and provide a reference for the utilization of the symmetry-protected topological surface states of this phase to realize novel topological phenomena and develop future device applications.

\bigskip
\hspace{0.5cm}
\textbf{Acknowledgements}
\newline


The work at Princeton and synchrotron x-ray-based measurements
are supported by the Office of Basic Energy
Sciences, US Department of Energy (DOE) Grant No. DEFG-
02-40105ER46200 and partial instrumentation support
provided by the Gordon and Betty Moore Foundations EPiQS
Initiative through Grant GBMF4547 (M.Z.H.). The work at
Northeastern University is supported by the DOE, Office of
Science, Basic Energy Sciences Grant Number DE-FG02-
07ER46352, and benefited from Northeastern University's
Advanced Scientific Computation Center (ASCC) and the
NERSC supercomputing center through DOE Grant Number
DE-AC02-05CH11231.
The use of Synchrotron Radiation Center (SRC) was supported by NSF DMR-0537588 under the external user agreement.
H.L. acknowledges the Singapore National Research Foundation for the support under NRF Award No. NRF-NRFF2013-03. 
T.D. acknowledges support of NSF IR/D program.
M.N. at LANL acknowledges  support from LANL LDRD program. 
We thank Sung-Kwan Mo, Alexi Fedorov and Makoto Hashimoto for beamline assistance at the LBNL and the SSRL.
 M.Z.H. acknowledges Visiting Scientist support from LBNL and additional support from DOE/BES and the A. P. Sloan Foundation.

\*Correspondence and requests for materials should be addressed to M.N. (Email: mneupane@lanl.gov) and M.Z.H. (Email: mzhasan@princeton.edu).

\end{document}